\documentclass[twocolumn,preprintnumbers,showpacs,nofootinbib]{revtex4-1}
 \usepackage{amssymb,amsmath,amsfonts}

\usepackage{graphicx}
\usepackage{setspace}
\usepackage{multirow}

\usepackage{color}

\newcommand{\mgp}{m_{\gamma'}}
\newcommand{\gp}{{\gamma'}}
\newcommand{\hp}{\ensuremath{\gamma' \ }}
\newcommand{\x}{\ensuremath{x_e}}
\newcommand{\xo}{\ensuremath{x_0}}

\begin{document}

\title{New limits on hidden photons from past electron beam dumps}

\author{Sarah Andreas}
\email{sarah.andreas@desy.de}
\author{Carsten Niebuhr}
\email{carsten.niebuhr@desy.de}
\author{Andreas Ringwald}
\email{andreas.ringwald@desy.de}
\affiliation{Deutsches Elektronen-Synchrotron DESY, 22607
Hamburg, Notkestrasse 85, Germany}

\begin{abstract}
Hidden sectors with light extra U(1) gauge bosons, so-called
hidden photons, have recently attracted some attention because
they are a common feature of physics beyond the Standard Model
like string theory and supersymmetry and additionally are
phenomenologically of great interest regarding recent
astrophysical observations. The hidden photon is already
constrained by various laboratory experiments and presently
searched for in running as well as upcoming experiments. We
summarize the current status of limits on hidden photons from
past electron beam dump experiments including two new limits
from such experiments at the High Energy Accelerator Research 
Organization in Japan (KEK) and the Laboratoire de l'accel\'{e}rateur lin\'{e}aire (LAL, Orsay) 
that have so far not been considered. All our limits take into account the
experimental acceptances obtained from Monte Carlo simulations.

\end{abstract}


\preprint{DESY 12-054}

\maketitle

\section{Motivation}
\label{Motivation} \reversemarginpar

Light extra U(1) gauge bosons appear very naturally in
well-motivated extensions of the standard model (SM). Even if
the paradigm of a highly symmetric high energy theory is to
unify particles into representations of a large-rank local
gauge group, the phenomenological fact that at low energies
these large gauge symmetries are seemingly broken possibly
leaves us with the existence of many lower rank symmetries. In
particular U(1)s are potentially most numerous since they are
the lowest-rank local symmetries. Moreover, some of these U(1)s
may be hidden, because the SM particles are not charged under
them, and therefore escaped detection until now. Most notably,
such hidden sectors often occur in supersymmetric extensions or
superstring embeddings of the
SM~\cite{Abel:2008ai,Goodsell:2009xc,Goodsell:2010ie,Cicoli:2011yh,Williams:2011qb}.

On general grounds, the dominant interaction of hidden U(1)
gauge bosons $\gamma'$ (hidden sector photons, short hidden
photons) with the SM at low energies generically appears
already at the dimension four level through kinetic mixing with
the ordinary photon~\cite{Holdom:1985ag}. Correspondingly, the
leading terms of the low energy effective Lagrangian of the
minimal hidden U(1) extension of the SM read,
\begin{equation}
{\cal L}_{\rm eff}~=~{\cal L}_{\rm SM} - \frac{1}{4}F_{\mu \nu}^\prime F^{\prime\,{\mu \nu}} +
\frac{m_{\gamma^{\prime}}^2}{2}  A^\prime_{\mu} A^{\prime\,\mu}
-\frac{\chi}{2} F^\prime_{\mu \nu} F^{\mu \nu} ,
\nonumber
\end{equation}
where $F^\prime_{\mu \nu}$ is the field strength of the hidden
gauge field $A_\mu^\prime$, $m_{\gamma^\prime}$ the mass of the
hidden photon, and $F_{\mu \nu}$ the field strength of the
ordinary electromagnetic gauge field.

The hidden photon mass $m_{\gamma^\prime}$ and the kinetic
mixing parameter $\chi$ have to be determined either
theoretically, by specifying an ultraviolet completion of the
theory, or phenomenologically, by comparing with observations.
In this context, two very interesting mass regions have been
identified recently:
\begin{itemize}
\item[i)] $m_{\gamma^\prime}\sim $~meV: Hidden photons in
    this mass range may explain the $\sim 2 \, \sigma$
    excess of dark radiation in the
    universe~\cite{Jaeckel:2008fi}, beyond the one from
    ordinary photons and neutrinos, reported by recent
    global cosmological
    analyses~\cite{Komatsu:2010fb,Dunkley:2010ge}. This
    possibility can be tested decisively in the next
    generation of light-shining-through-a-wall
    experiments~\cite{Redondo:2010dp}.
\item[ii)] $m_{\gamma^\prime}\sim $~GeV: Hidden photons in
    this mass range may explain the observed $\sim
    3\,\sigma$ deviation of the anomalous magnetic moment
    of the muon~\cite{Bennett:2006fi} from the value
    expected in the SM~\cite{Pospelov:2008zw}. Moreover,
    they might explain possible terrestrial and cosmic ray
    dark matter anomalies --- notably the possible direct
    detection of dark matter by
    DAMA~\cite{Bernabei:2008yi},
    CoGeNT~\cite{Aalseth:2010vx,Aalseth:2011wp} and
    CRESST~\cite{Angloher:2011uu,Stodolsky:2012wf}, in
    contrast to its nonobservation in
    CDMS~\cite{Ahmed:2009zw} and
    XENON~\cite{Aprile:2011hi}, and the observations of an
    excess in cosmic ray electrons and/or positrons
    observed by PAMELA~\cite{Adriani:2008zr} and
    Fermi~\cite{Abdo:2009zk} --- if dark matter resides in
    the hidden sector too and is charged under the hidden
    U(1)~\cite{ArkaniHamed:2008qn,Pospelov:2008jd,Feldman:2008xs,Morrissey:2009ur,Mambrini:2010dq,Mambrini:2011dw,Andreas:2011in}.
    This possibility can be tested seriously with new
    accelerator based
    experiments~\cite{Batell:2009yf,Essig:2009nc,Reece:2009un},
    especially with new beam dump and fixed target
    experiments exploiting high intensity
    electron~\cite{Bjorken:2009mm,Freytsis:2009bh} and
    proton beams~\cite{Batell:2009di,Essig:2010gu}.
\end{itemize}

Motivated by this strong physics case, a number of electron
beam dump and fixed target experiments to search for GeV scale
hidden photons (\textit{dark forces}) have been proposed or even
taken first
data~\cite{Essig:2010xa,Merkel:2011ze,Abrahamyan:2011gv}. In
this context it is very important to analyze also results from
past electron beam dump experiments in terms of hidden photons
--- a task which was accomplished quite exhaustively in
Ref.~\cite{Bjorken:2009mm}. However, in this paper two
experiments have not been considered: {\em i)} a beam dump
experiment searching for neutral penetrating particles
exploiting an electron linear accelerator at the High Energy Accelerator Research 
Organization in Japan (KEK)~\cite{Konaka:1986cb} and {\em ii)} a beam dump experiment
exploiting the Orsay Linac originally analyzed in terms of
production and late decay of scalars (\textit{Higgs}) and
pseudoscalars (\textit{axions})~\cite{Davier:1989wz}\footnote{In Ref. \cite{Bouchiat:2004sp} it had already been
suggested that the electron beam dump experiment at Orsay could be
used to constrain the more general $U$-boson for which another limit
from proton beam dumps was obtained in Ref. \cite{Fayet:1980rr}.}. In this
paper, we derive the corresponding bounds on hidden photons,
which exceed the bounds previously established by other
electron beam dump experiments in a certain region of the
parameter space.

\section{\boldmath $\gamma^\prime$ in Electron Beam Dumps}

In this section, we summarize the relevant formula (based
on Ref.~\cite{Bjorken:2009mm}) and computational steps necessary to
determine the expected signatures of a hidden photon $\gp$ in a
beam dump experiment and to deduce the limits on its mass
$m_\gp$ and kinetic mixing $\chi$ set by different experiments.

\subsection{$\gp$ production in bremsstrahlung}
\label{sec-crosssection} Hidden photons are generated in
electron (or positron) collisions on a fixed target by a
process analogous to ordinary photon bremsstrahlung. For an
incoming electron with energy $E_e$ the corresponding
differential cross section in the range
\begin{equation}
m_e \ll \mgp \ll E_e \quad \mathrm{and} \quad
x_e \theta_\gp^2 \ll 1
\end{equation}
is given in Ref.~\cite{Bjorken:2009mm} Eq.~(A12) by
\begin{align}
& \frac{d\sigma}{d\x \ d\cos\theta_\gp} = ~ 8 \alpha^3 \chi^2 E_e^2 \x \ \xi
\sqrt{1 - \frac{\mgp^2}{E_e^2}} \label{eq-dsdxdtheta} \\
& ~~~~\left[ \frac{1-\x+\frac{\x^2}{2}}{U^2}  +  \frac{(1-\x)^2 \mgp^4}{U^4} -  \frac{(1-\x) \x \mgp^2}{U^3} \right] \nonumber
\label{eq-dsdxdtheta}
\end{align}
where $\x=E_\gp/E_e$ is the fraction of the incoming electron's
energy carried by the hidden photon, $\theta_\gp$ is the lab
frame angle between emitted \hp and incoming electron and $Z$
and $A$ are atomic number and mass number of the nucleus in the
target. The effective flux of photons $\xi$ is given by
\begin{equation}
\xi(E_e, \mgp, Z, A) ~=~ \int_{t_\mathrm{min}}^{t_\mathrm{max}} dt \ \frac{t - t_\mathrm{min}}{t^2} \ G_2(t) \ ,
\label{eq-xi}
\end{equation}
where $t_\mathrm{min}= ( \mgp^2/2 E_e )^2$, $t_\mathrm{max}=
\mgp^2$ and the electric form factor $G_2(t)$ defined
in Ref.~\cite{Bjorken:2009mm} consists of an elastic and an
inelastic contribution both of which depend on the atomic
number $Z$ and mass $A$. The function $U$ describes the
virtuality of the intermediate electron in initial-state
bremsstrahlung and is given by
\begin{equation}
U(\x,E_e,\mgp,\theta_\gp) = E_e^2 \ \x \ \theta_\gp^2 + \mgp^2 \frac{1-\x}{\x} + m_e^2 \ \x \ . \label{eq-U}
\end{equation}
Integrating Eq.~(\ref{eq-dsdxdtheta}) over the emission angle
$\theta_\gp$ of the hidden photon from $0$ to some maximum
angle $\theta_\mathrm{max}$ set by the geometry of the
experiment (for the experiments under consideration
$\theta_\mathrm{max}<0.5\ \mathrm{rad})$ we
obtain~\footnote{Note that this expression includes a factor
1/2 which has erroneously been omitted
in Ref.~\cite{Bjorken:2009mm}.}
\begin{equation}
\frac{d\sigma}{d\x} ~=~ 4 \alpha^3 \chi^2 \ \xi \ \sqrt{1 -
\frac{\mgp^2}{E_e^2}} \ \frac{1 - \x + \frac{\x^2}{3}}{\mgp^2
\frac{1-\x}{\x} + m_e^2 \x} \label{eq-dsdxapprme} \ .
\end{equation}

\subsection{Number of expected events behind a beam dump}
For a beam dump experiment with an electron beam of energy
$E_0$ incident on a target (cf. Fig.~\ref{fig-ExpSketch}) one
has to take into account that the initial energy of the
electrons in the beam becomes degraded as they pass through the
target and interact with the material. This is described by the
energy distribution of the electrons after passing through a
medium of $t$ radiation length which according
to Tsai~\cite{Tsai:1986tx} is roughly given by
\begin{equation}
I_e (E_0,E_e,t) = \frac{1}{E_0} \
\frac{\left[\ln\left(\frac{E_0}{E_e}\right)\right]^{bt-1}}{\Gamma(bt)} \ , \label{eq-Ie}
\end{equation}
where $E_0$ is the initial monochromatic electron beam energy
at $t=0$, $\Gamma$ is the Gamma function and $b=\frac{4}{3}$.
The bremsstrahlung cross section from the previous subsection
which depends on the energy $E_e$ of the electrons therefore
has to be convoluted with this energy distribution and
integrated over the length $L_\mathrm{sh}$ of the target plus
shield. Together with Eq.~(\ref{eq-Ie}), the total number of
hidden photons with an energy $E_\gp \equiv \xo E_0$ that are
produced in the target via bremsstrahlung off the electron beam
and that decay at a distance $z$ behind the front edge of the
target is then given by
\begin{align}
\frac{dN}{d\xo \ dz} = \ & N_e \frac{N_0 X_0}{A} \int_{E_\gp+m_e}^{E_0} \! dE_e \int_0^T \! dt
\Bigg[I_e(E_0,E_e,t) \nonumber \\
& ~~ \frac{E_0}{E_e} \ \left. \frac{d\sigma}{d\x} \right|_{\x=\frac{E_\gp}{E_e}} \frac{dP(z-\frac{X_0}{\rho}t)}{dz} \Bigg] \ , \label{eq-dNdxdz}
\end{align}
where $N_e$ and $E_0$ are the number and energy of the incident
electrons, respectively, $N_0 \simeq 6 \times 10^{23} \;
\mathrm{mole}^{-1}$ is Avogadro's number, $\rho \
[\mathrm{g/cm}^3]$ and $X_0 \ [\mathrm{g/cm}^2]$ are the
density and unit radiation length of the target material,
respectively, and $T \equiv \rho L_\mathrm{sh}/X_0$ is the
length $L_\mathrm{sh}$ of target plus shield in units of
radiation length. The differential cross section $d\sigma/d\x$
discussed in Sec.~\ref{sec-crosssection} is given in
Eq.~(\ref{eq-dsdxapprme}). The differential decay probability
$dP/dz$ is defined as
\begin{equation}
\frac{dP(l)}{dl} = \frac{1}{l_\gp} \ e^{-l/l_\gp} \ , \label{eq-dPdl}
\end{equation}
where $l_\gp$ is the decay length of the hidden photon $l_\gp =
\gamma \tau_\gp = \frac{E_\gp}{\mgp} \ \frac{1}{\Gamma_\gp}$.
For the mass range of interest, the total decay width
$\Gamma_\gp$ is given by
\begin{equation}
\Gamma_\gp ~=~ \Gamma_{\gp \rightarrow e^+ e^-} \ + \ \Gamma_{\gp \rightarrow \mu^+ \mu^-} \left[ 1 +
R(\mgp) \right] \ , \label{eq-Gammagp}
\end{equation}
where the second term is only present for $m_\gp \geq 2 m_\mu$,
the partial decay width into leptons is given
by~\cite{Pospelov:2008zw}
\begin{equation}
\Gamma_{\gp \rightarrow l^+ l^-} = \frac{\alpha \chi^2}{3} \mgp
\left( 1 + 2 \frac{m_l^2}{\mgp^2} \right) \sqrt{1 - 4 \frac{m_l^2}{\mgp^2}} \ ,
\end{equation}
and $R(\sqrt{s})$ is defined as the energy dependent ratio
$\frac{\sigma(e^+e^- \rightarrow \mathrm{hadrons}, \
\sqrt{s})}{\sigma(e^+e^- \rightarrow \mu^+ \mu^-, \ \sqrt{s})}$
taken from Ref.~\cite{Amsler:2008zzb}.

\subsection{Special case: Thick target beam dump experiment}
\begin{figure}[htb!]
\begin{center}
\includegraphics[width=\columnwidth]{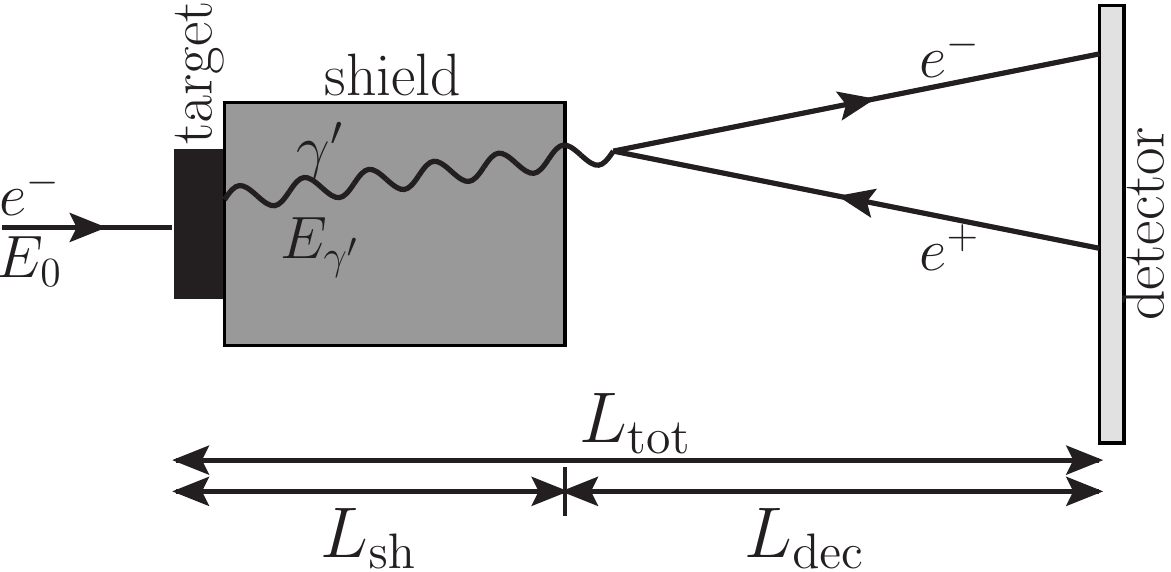}
\caption{Sketch of the setup of an electron beam dump experiment illustrating the definitions
of the lengths $L_\mathrm{sh}$, $L_\mathrm{dec}$ and $L_\mathrm{tot}$ used in the text.
An incoming electron beam of energy~$E_0$ hits the target and produces in bremsstrahlung a hidden photon $\gp$
with energy~$E_\gp$ that decays behind the shield e.g. into $e^+e^-$ which can then be observed
in the detector. } \label{fig-ExpSketch}
\end{center}
\end{figure}

In the case of a thick target experiment, which we are
interested in, most of the hidden photon production takes place
within the first radiation length so that the $t$ dependence in
the $\gp$ decay probability can be neglected and
Eq.~(\ref{eq-dNdxdz}) simplifies to
\begin{align}
\frac{\ dN}{d\xo \ dz} \simeq \ & N_e \frac{N_0 X_0}{A} \int_{E_\gp+m_e}^{E_0} \! dE_e \int_0^T \! dt
\Bigg[I_e(E_0,E_e,t) \nonumber \\
& ~~ \frac{E_0}{E_e} \ \left. \frac{d\sigma}{d\x} \right|_{\x=\frac{E_\gp}{E_e}} \frac{dP(z)}{dz} \Bigg] \ . \label{eq-dNdxdzthick}
\end{align}
After carrying out the integration over $z$ from
$L_\mathrm{sh}$ to \mbox{$L_\mathrm{tot} \equiv L_\mathrm{sh} +
L_\mathrm{dec}$}, where $L_\mathrm{sh}$ is the length of target
plus shield and $L_\mathrm{dec}$ the length of the decay
region, as sketched in Fig.~\ref{fig-ExpSketch}, this becomes
\begin{align}
\frac{dN}{d\xo} \simeq \ & N_e \ \frac{N_0 X_0}{A} \ \int_{E_\gp+m_e}^{E_0} \! dE_e \ \int_0^T \! dt \
\Bigg[ I_e(E_0,E_e,t) \nonumber \\
& ~ \frac{E_0}{E_e} \ \left. \frac{d\sigma}{d\x} \right|_{\x=\frac{E_\gp}{E_e}}^{\vphantom{\int}}
e^{-L_\mathrm{sh}/l_\gp} \left( 1 - e^{-L_\mathrm{dec}/l_\gp} \right) \Bigg] \ .
\end{align}

The total number of events behind the dump resulting from the
decay of the hidden photon is then given by
\begin{align}
N \simeq \ & N_e \frac{N_0 X_0}{A} \int_{\mgp}^{E_0-m_e} \! dE_\gp \int_{E_\gp+m_e}^{E_0} \! dE_e \int_0^T \! dt \nonumber \\
& \Bigg[I_e(E_0,E_e,t) \ \frac{1}{E_e} \left.
\frac{d\sigma}{d\x} \right|_{\x=\frac{E_\gp}{E_e}} \nonumber \\
&~~ e^{-L_\mathrm{sh}/l_\gp} \left( 1 - e^{-L_\mathrm{dec}/l_\gp}
\right) \Bigg] \ \mathrm{BR}_\mathrm{detect}\ . \label{eq-Nevents}
\end{align}
where $\mathrm{BR}_\mathrm{detect}$ is the branching ratio into
those decay products that the detector is sensitive to, i.e.
electrons or muons or both.

\renewcommand{\arraystretch}{1.8}
\setlength{\tabcolsep}{9pt}
\begin{table*}[!t]
\begin{center}
\begin{tabular}{l c r r r r r c c }
\multirow{2}{*}{Experiment} & \multirow{2}{*}{target} & \multicolumn{1}{c}{$E_0 $} & \multicolumn{2}{c}{$N_{\mathrm{el}}$}
& \multicolumn{1}{c}{$L_{\mathrm{sh}}$} & \multicolumn{1}{c}{$L_{\mathrm{dec}}$}
& \multirow{2}{*}{$N_{\mathrm{obs}}$} & \multirow{2}{*}{$N_{\mathrm{95\% up}}$} \vspace{-0.2cm}\\
&  & \multicolumn{1}{c}{$[\mathrm{GeV}]$} & \multicolumn{1}{c}{electrons} & \multicolumn{1}{c}{Coulomb}
& \multicolumn{1}{c}{$[\mathrm{m}]$} &  \multicolumn{1}{c}{$[\mathrm{m}]$}  &  &  \\
\hline
E141~\cite{Riordan:1987aw} & W & 9  $~$ & 2$\times 10^{15}$ & 0.32 mC & 0.12  & 35   & $1126^{+1312}_{-1126}$ & 3419 \\
E137~\cite{Bjorken:1988as} & Al & 20  $~$ & 1.87$\times 10^{20}$ & 30 \hphantom{m}C & 179  & 204   & 0 & 3 \\
E774~\cite{Bross:1989mp} & W & 275  $~$ & 5.2$\times 10^{9\hphantom{1}}$ & 0.83 $\ $nC & 0.3  & 2   & $0^{+9}_{-0}$ & 18  \\
KEK~\cite{Konaka:1986cb} & W & 2.5 $~$  & 1.69$\times 10^{17}$ & 27 mC & 2.4   & 2.2   & 0 & 3 \\
Orsay~\cite{Davier:1989wz} & W & 1.6  $~$ & 2$\times 10^{16}$ & 3.2 mC & 1  & 2   & 0 & 3 \\
\end{tabular}\vspace{0.3cm}
\caption{Overview of the different beam dump experiments
analyzed in this work and their specifications. The number of
observed events $N_{\mathrm{obs}}$ have directly been extracted
from the experiment's papers and differ in the case of E141 and
E137 slightly from the estimates used in Ref.~\cite{Bjorken:2009mm}
as do the corresponding $95 \%$ C.L. values.\label{tab-Exp}
}\end{center}
\end{table*}

\subsection{Acceptance of different experiments} \label{sec-acceptance}
Up to now we have not taken into account that depending on the
angle under which the final decay products are emitted and the
geometry of the detector, not all events computed according to
Eq.~(\ref{eq-Nevents}) are actually seen by the detector. With
the use of \textsc{MadGraph}~\cite{Alwall:2011uj,EST:MC} we generated
for the different experiments (see Table~\ref{tab-Exp}) Monte
Carlo simulations of the hidden photon's production in
bremsstrahlung followed by its decay into $e^+e^-$. Comparing
the thereby obtained decay angles with the geometrical setup of
the experiment an acceptance specific for each experiment can
be determined. Repeating this procedure for every experiment
along the rough exclusion contour obtained using
Eq.~(\ref{eq-Nevents}) we can rescale the limit with the proper
acceptance to get the final exclusion region. In the cases
where the acceptances have been given in the experiment's
paper, we compared and found them in reasonable agreement with
the results of our Monte Carlo simulations.

\section{Electron Beam Dump Experiments}

An overview of the different electron beam dump experiments and
their properties is shown in Table~\ref{tab-Exp}.
In Ref.~\cite{Bjorken:2009mm}, the limits set by the E141 and E137 
experiments at the Stanford Linear Accelerator Center (SLAC) 
as well as the Fermilab E774 experiment have
already been analyzed. In the present paper, we extend their
analysis by two experiments that so far have not been
considered: one electron beam dump experiment at KEK in
Japan~\cite{Konaka:1986cb} and one at the Orsay Linac in
France~\cite{Davier:1989wz}. In addition, our analysis includes
the experimental acceptances obtained from Monte Carlo
simulations with \textsc{MadGraph} in the determination of the limits
for all experiments, as described in Sec.~\ref{sec-acceptance}.

\begin{figure}[b!]
\begin{center}
\includegraphics[width=\columnwidth]{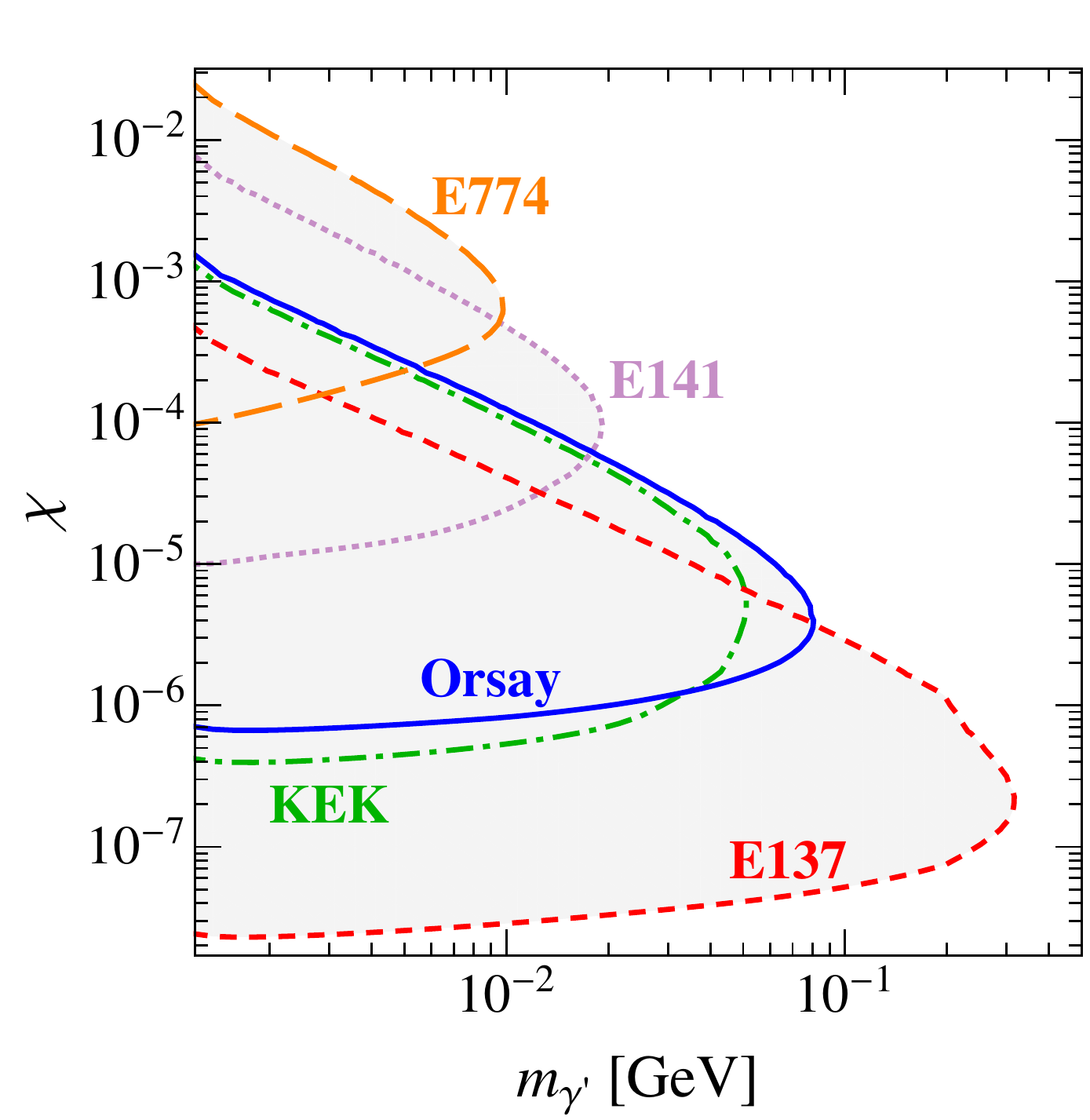}
\caption{Limits on the hidden photon mass $m_\gp$ and kinetic
mixing $\chi$ from different electron beam dump experiments. In
addition to the limits from E141 (magenta
dotted line), E137 (red dashed line) and E774 (orange
long-dashed line) presented already in Ref.~\cite{Bjorken:2009mm},
the regions labeled KEK (green dash-dotted line) and Orsay
(blue solid line) have been excluded in the present work.}
\label{fig-limits}
\end{center}
\end{figure}

For the \textbf{SLAC E141} experiment~\cite{Riordan:1987aw}, we extract from Fig.~1c for
$x\geq0.7$ a total of 1126$\pm$1312 events, which corresponds to a 95\% C.L. upper limit
of $N_{95\%\mathrm{up}} = 3419$ events. The appropriate
exclusion contour shown in Fig.~\ref{fig-limits} takes into
account the acceptance from \textsc{MadGraph}.

As the \textbf{SLAC E137} experiment reported
in Ref.~\cite{Bjorken:1988as} that no candidate events were
observed in their search for axionlike particles, the 95\%
C.L. upper limit is given by \mbox{$N_{95\%\mathrm{up}} = 3$}
events. Together with the acceptance we then find the exclusion
contour presented in Fig.~\ref{fig-limits}.

For the \textbf{Fermilab E774} experiment we find a total of
zero events with excess multiplicity 2 from Fig.~4c
in Ref.~\cite{Bross:1989mp}. Resulting from a substraction of the
background from the original spectrum, the statistical error of
$\sqrt{89}$ is dominated by the total number of events in
Fig.~4b. The acceptance corrected 95\% C.L. upper limit of
\mbox{$N_{95\%\mathrm{up}} = 18$} events leads to the exclusion
contour in Fig.~\ref{fig-limits}.

In the electron beam dump experiment at
\textbf{KEK}~\cite{Konaka:1986cb} no signal was observed in
their search for axionlike particles. The corresponding 95\%
C.L. upper limit $N_{95\%\mathrm{up}}$ of 3 events together
with the acceptance leads for hidden photons to the exclusion
contour presented in Fig.~\ref{fig-limits}.

The electron beam dump experiment in
\textbf{Orsay}~\cite{Davier:1989wz} also found no positive
signal when looking for light Higgs bosons. This translates to
a 95\% C.L. upper limit $N_{95\%\mathrm{up}}$ of 3 events on
hidden photons. Considering the experiment's acceptance we find
the exclusion contour shown in Fig.~\ref{fig-limits}.

\section{Discussion}

As presented in Fig.~\ref{fig-limits}, the experiments at KEK
and in Orsay were found to exclude a similar region of the
parameter space which so far has not been constrained by any
other electron beam dump experiment. Our limits from the
previously analyzed experiments at SLAC and Fermilab are
comparable to those derived in Ref.~\cite{Bjorken:2009mm} but are
generally slightly weaker because of the factor 1/2 discrepancy
in Eq.~(\ref{eq-dsdxapprme}), the fact that our Monte Carlo
simulations yield somewhat different experimental acceptances
and due to a little different numbers of events
$N_{\mathrm{95\% up}}$ used for our $95 \%$ C.L. contours.

\begin{figure}[htb!]
\begin{center}
\includegraphics[width=\columnwidth]{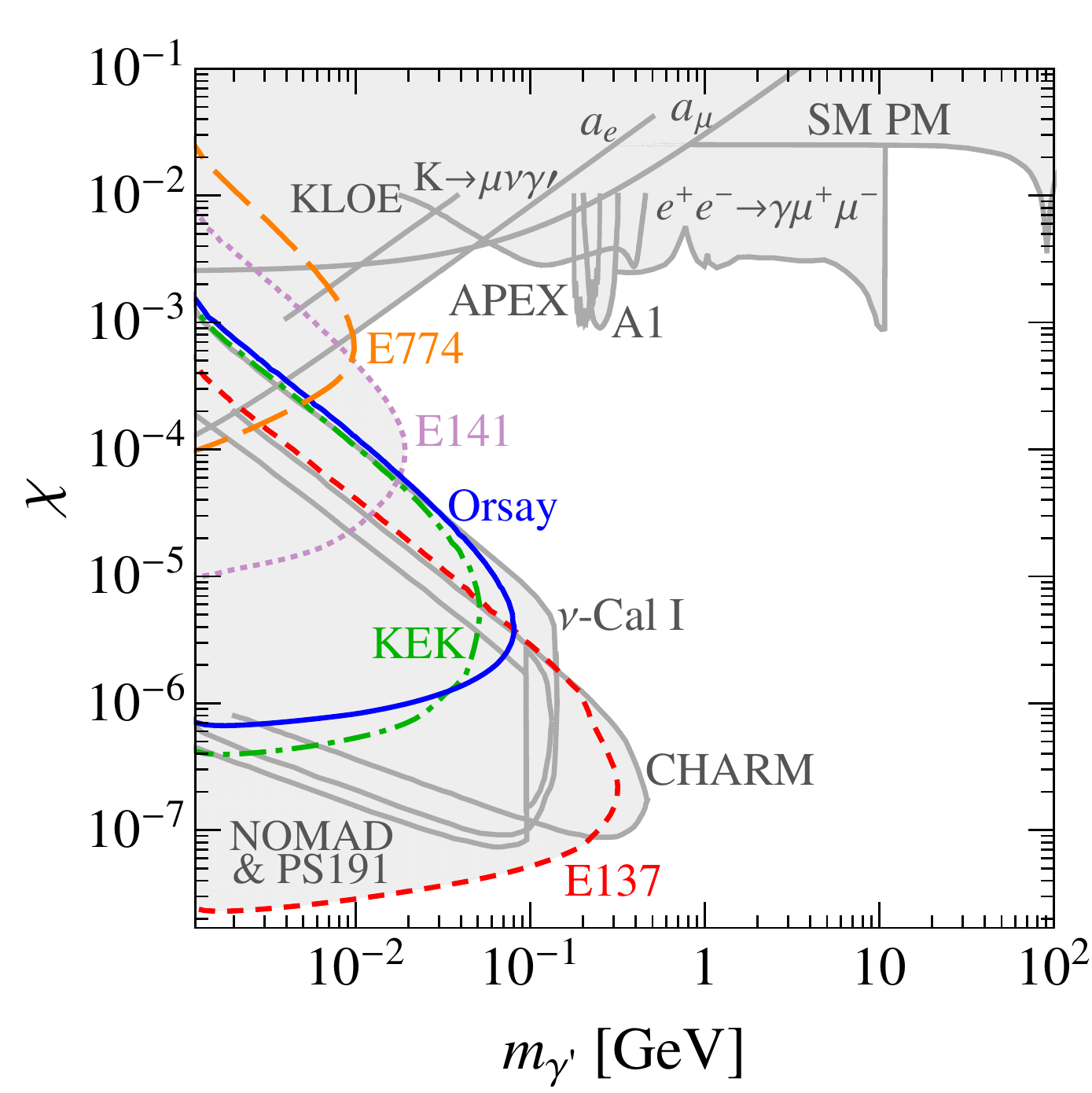}
\caption{Collection of all current limits on hidden photons:
from the electron beam dump experiments of the present work
(colored lines, cf. Fig.~\ref{fig-limits}, all
other limits as gray lines), Standard Model precision
measurements, muon and electron anomalous magnetic moment,
a reinterpretation of the BaBar search $e^+ e^- \rightarrow \gamma \mu^+ \mu^-$ for pseudoscalars, the electron fixed target experiments A1 and APEX, the $\nu$-Cal I experiment at the Serpukhov proton beam dump, the KLOE experiment, the neutrino experiments NOMAD, PS191 and CHARM, and from the Kaon decay $ K \rightarrow \mu \nu \gamma' $, cf. text for
details.} \label{fig-limits-all}
\end{center}
\end{figure}

Besides the limits from electron beam dump experiments
discussed in this article, there are various other constraints
on the hidden photon mass and kinetic mixing which we briefly
summarize in the following. Their comparison to the electron
beam dump experiments from Fig.~\ref{fig-limits} is shown in
Fig.~\ref{fig-limits-all}.\vspace{-0.04cm}
\vspace{-0.04cm}

In Ref.~\cite{Hook:2010tw}, SM precision measurements were found to
exclude large values of hidden photon mass and kinetic mixing
as indicated in Fig.~\ref{fig-limits-all} by the label ``SM
PM''. Furthermore, as presented in Ref.~\cite{Pospelov:2008zw} the
muon and electron anomalous magnetic moment receiving a
one-loop contribution from the hidden photon place additional
constraints labeled ``$a_\mu$'' and ``$a_e$'' respectively; the latter was updated~\cite{Davoudiasl:2012ig,Endo:2012hp} while this work has been completed.
The reinterpretation of the BaBar search for a pseudoscalar
around the $\Upsilon(3S)$ resonance~\cite{Aubert:2009au} in the
process $e^+ e^- \to \gamma \mu^+ \mu^-$ was used
in Ref.~\cite{Reece:2009un,Essig:2010xa,Hook:2010tw,Echenard:2012iq} to derive a
limit on hidden photons. The two fixed target experiments A1 at
MAMI in Mainz~\cite{Merkel:2011ze} and APEX at
JLab~\cite{Abrahamyan:2011gv} both searching for hidden photons
behind a thin target from bremsstrahlung off an electron beam
started recently and were already able to set first new
limits. Reanalyzing proton beam dump data from the $\nu$-Cal~I experiment at the
U70 accelerator at IHEP Serpukhov a region overlapping with the
one of KEK and Orsay has been excluded
in Ref.~\cite{Blumlein:2011mv}. The \mbox{KLOE-2}
experiment~\cite{Archilli:2011zc} at the Frascati DA$\phi$NE
$\phi$-factory uses $e^+e^-$ collisions to place further
constraints. 
Very recently the production of hidden photons in the radiative decays of neutral pseudoscalar mesons, generated by a proton beam in neutrino experiments at CERN, has been constrained with NOMAD and PS191~\cite{Gninenko:2011uv} in the decay of $\pi^0$ and CHARM~\cite{Gninenko:2012eq} in the one of $\eta$ and $\eta'$.  
While this work was completed a new limit from the Kaon decay $ K \rightarrow \mu \nu \gamma' $ was derived~\cite{Beranek:2012ey} which would have improved the previous $a_e$ bound but is not competitive with the updated one.

An up-to-date overview of all current constraints on the mass
$m_\gp$ and kinetic mixing $\chi$ of the hidden photon from
various searches including the electron beam dump experiments
presented in this work is shown in Fig.~\ref{fig-limits-all}.
Despite the large number of constraints a broad region of the
parameter space remains open and is partly going to be tested
in currently already
running~\cite{Merkel:2011ze,Abrahamyan:2011gv,Essig:2010xa} and
planned future
experiments~\cite{HPS,Wojtsekhowski:2009vz,Freytsis:2009bh}, see Ref.~\cite{Hewett:2012ns} for an overview.\\

\vspace{-1cm}
\begin{acknowledgments}
We would like to thank  Yuri Soloviev for pointing out
Ref.~\cite{Konaka:1986cb} and Jonathan Jacobsohn for discussions at an early stage of this work. We are also grateful that Rouven
Essig, Philip Schuster and Natalia Toro provided us with their
\textsc{MadGraph} code for hidden photons which allowed us to simulate
the acceptances of the different beam dump experiments. Special
thanks to Rouven Essig for helpful discussions regarding the
factor 1/2 discrepancy in the cross section.
\end{acknowledgments}

\end{document}